# Localization of light on a cone: theoretical evidence and experimental demonstration for an optical fiber

M. Sumetsky
OFS Laboratories, 19 Schoolhouse Road, Somerset, NJ 08807
*Email: sumetski@ofsoptics.com*

The classical motion at a conical surface is bounded at one (narrower) side of the cone and unbounded at the other. However, it is shown here that a dielectric cone with a small half-angle $\gamma$ can perform as a high Q-factor optical microresonator which completely confines light. The theory of the discovered localized conical states is in excellent agreement with experimental data. It provides both a unique approach for extremely accurate local characterization of optical fibers (which usually have $\gamma \sim 10^{-5}$ or less) and a new paradigm in the field of high Q-factor resonators.

## 1. Introduction

Usually, a high Q-factor of a whispering gallery mode (WGM) in dielectric microresonators (e.g., in microspheres and microtoroids [1,2]) is a result of strong WGM localization due to the total internal reflection from the resonator surface and/or classical conservation laws, which confine the mode between two caustics separating the classically allowed and classically forbidden regions [3,4]. It is commonly accepted that the absence of caustics, strongly reflecting material interfaces, or periodic perturbations makes the wave motion unbounded and leads to the disappearance of the high Q-factor resonances. For this reason, it could be expected that all the WGMs in a lossless dielectric cone illustrated in Fig. 1 are delocalized. In fact, the classical motion at the conical surface is bounded on the narrower side of the cone and is unbounded on its wider side, so that any geodesic (classical ray) propagating at the conical surface eventually moves off to infinity (Fig. 1(a)). In contrast, it is shown below that, for a cone with a small half-angle $\gamma$, a wave beam launched normally to the cone axis (using, e.g., an optical microfiber [5]) can be completely localized (Fig. 1(b)).

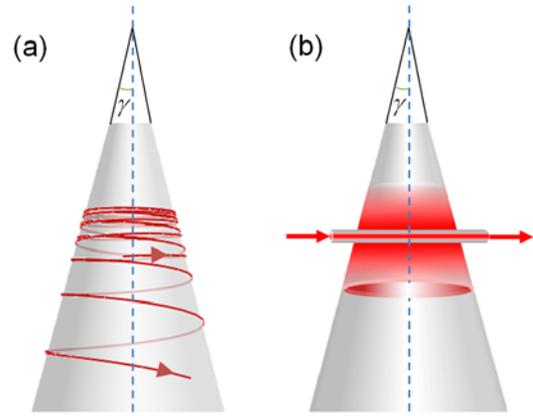

Fig. 1 (a) – Illustration of a geodesic propagating along the conical surface. (b) – Illustration of a localized WGM launched by a microfiber.

The discovered localized conical modes are common for conventional optical fibers which usually have $\gamma \sim 10^{-5}$ or less. It is found that the transmission resonance shape of a conical WGM exhibits asymmetric Airy-type oscillations, which allow one to determine the local slope $\gamma$ of a slightly nonuniform microcylinder (e.g., of an optical fiber) from a single measurement. As cone half-angle $\gamma$ decreases, the size of the localized mode grows very slowly, as $\gamma^{-1/3}$. The developed theory, applied to the investigation of the local slope of an optical fiber, is in excellent agreement with the experimental data. It provides both a unique approach for extremely accurate local characterization of optical fibers (which usually have $\gamma \sim 10^{-5}$ or less) and a new paradigm in the field of high Q-factor resonators.

## 2. Theory

It is assumed that a light beam propagating close to the conical surface is launched by a microfiber waveguide which touches the cone surface normally to its axis (Fig. 1(b)). The geometry of light propagation along the conical surface can be better visualized by unfolding it into a planar surface as shown in Fig. 2 [6]. Then, all the conical geodesics experiencing multiple turns are transformed into straight lines. Similarly, a light beam propagating along a curved geodesic at the cone is transformed into a beam which diffracts in the vicinity of a straight line at the unfolded surface. For a small cone half-angle $\gamma$, the radial dependence of the propagating beam (i.e., dependence on the coordinate normal to the unfolded surface) remains unchanged and is omitted in further analysis.

A beam launched in the vicinity of the microfiber/cone contact point $\varphi = z = 0$ (here $\varphi$ is the azimuthal angle and $z$ is the fiber axial coordinate introduced in the inset of Fig. 2) can be expanded into a linear combination of planar Gaussian beams of order $n$ having the form $(is_0 + s)^{-n/2} H_n[\beta^{1/2} z(is_0 + s)^{-1/2}] \exp[i(\beta + i\alpha)s]$. Here $\beta$ and $\alpha$ are the propagation and attenuation constants, $H_n(x)$ is a Hermite polynomial, $s_0$ determines the beam waist at the launch point, and $s = s(\varphi, z)$ is the distance between the original point $\varphi = z = 0$ and point $(\varphi, z)$ calculated along the unfolded conical surface (Fig. 2). After a large number of turns $m$, (i.e., large distance $s$) the beams with $n > 0$ vanish as $s^{-(n+1)/2}$ and become negligible compared to the fundamental Gaussian beam with $n = 0$. In addition, for large $s$ the waist parameter $s_0$ can also be neglected. For a weak microfiber/cone coupling (i.e., in the strongly undercoupling regime

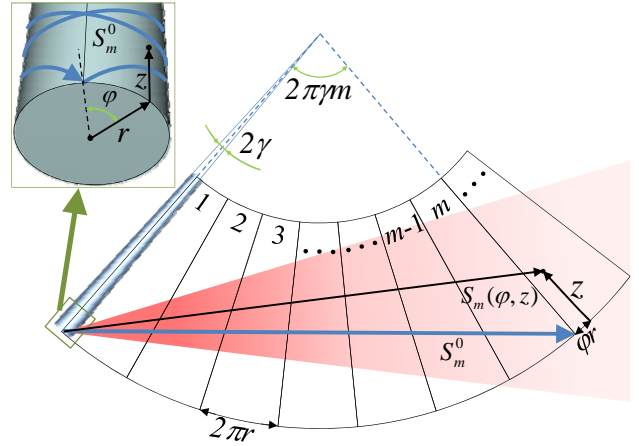

Fig. 2. Unfolded conical surface with a Gaussian beam propagating along this surface. The inset shows a cone segment with a curved geodesic $S_m^0$ which corresponds to the straight trajectory $S_m^0$ on the unfolded surface.

[7]), the resonant field at point $(\varphi, z)$ of the cone surface is found as the following superposition of fundamental Gaussian beams which are launched at point $\varphi = z = 0$ and make $m$ turns before approaching point $(\varphi, z)$:

$$\Psi(\varphi, z) \sim \sum_m S_m(0,0)^{-1/2} \exp\left[i(\beta + i\alpha) S_m(\varphi, z)\right] \quad (1)$$

Here $S_m(\varphi, z)$ is the distance between the launch point and point $(\varphi, z)$ calculated along the geodesic which connects these points after completing $m$ turns (Fig. 2). Eq. (1) is a semiclassical representation [8] for the considered WGMs. As follows from Fig. 2, for $2\pi m\gamma \ll 1$, we have

$$S_m(\varphi, z) \approx S_m^0 + \varphi r - \pi m \gamma z + z^2 / (2 S_m^0) \quad (2)$$

where $r$ is the local cone radius of the circumference $(\varphi, 0)$ and $S_m^0$ is the length of the geodesic crossing itself at the original point after $m$ turns (Fig. 2):

$$S_m^0 = \frac{2r}{\gamma} \sin(\pi \gamma m) \approx 2\pi r m - \frac{\pi^3}{3} \gamma^2 r m^3 \quad (3)$$

The resonance propagation constant $\beta_q$ can be defined by the quantization condition along the circumference $(\varphi, 0)$: $\beta_q = q / r$, where $q$ is a large integer. Assuming that the sum in Eq. (1) is determined by terms with large number $m \gg 1$ we replace it with an integral. Using Eq. (2) and (3) we get

$$\Psi(\varphi, z) \sim \exp(iq\varphi) \int_0^\infty \frac{dm}{m^{1/2}} \exp\left\{\pi i \left[2(\Delta\beta + i\alpha)r - \beta_q \gamma z\right] m - \frac{i\pi^3 \beta_q}{3} \gamma^2 r m^3 + \frac{i\beta_q z^2}{4\pi r m}\right\} \quad (4)$$

where $\Delta\beta = \beta - \beta_q$ is the deviation of the propagation constant. In this expression, the first term in the square brackets and the last term in the exponent correspond to the usual Gaussian beam propagating along the straight line. The terms proportional to $\gamma^2 m^3$ and $\gamma m$ following from Eqs. (2) and (3) characterize the curved geodesic and are responsible for the major effects described below. Eq. (4) is valid if the deviation of the propagation constant, $\Delta\beta$, attenuation, $\alpha$, and the cone slope, $\gamma$, are small, i.e., if $\Delta\beta, \alpha \ll (2\pi r)^{-1}$, and

$$\gamma \ll \pi^{-3/2}(\beta r)^{-1/2}. \tag{5}$$

For a conventional optical fiber of radius $r \sim 50\ \mu\text{m}$ and effective refractive index $n_r \sim 1.5$, at radiation wavelength $\lambda \sim 1.5\ \mu\text{m}$ we have $\beta = 2\pi n_r / \lambda \sim 6\ \mu\text{m}^{-1}$ and Eq. (5) is satisfied for $\gamma \ll 10^{-2}$.

If the microfiber/cone coupling is localized near $\varphi = z = 0$ then the resonant transmission power is found from Eq. (4) as $P = |1 - D - C\Psi(0,0)|^2$ where parameters $C$ and $D$ are constants in a vicinity of the resonance. For weak coupling considered here, $|D|, |C\Psi(0,0)| \ll 1$ so that

$$P \approx 1 - 2\operatorname{Re}(D) - 2\operatorname{Re}\left\{C\int_0^\infty \frac{dm}{m^{1/2}}\exp\left[2\pi r(i\Delta\beta - \alpha)m - \frac{i\pi^3 \beta_q}{3}\gamma^2 r m^3\right]\right\}. \tag{6}$$

Two alternatives are possible. For a relatively small cone angle, $\gamma \ll 5\alpha^{3/2}\beta^{-1/2} r$, the terms depending on $\gamma$ in the exponents of Eqs. (4) and (6) can be neglected and the conical resonator behaves as a uniform cylindrical resonator investigated in [9]. Alternatively, for

$$\gamma \gg 5\alpha^{3/2}\beta^{-1/2} r \tag{7}$$

the effect of loss is suppressed by the slope value $\gamma$ and the term $\sim \alpha$ in Eqs. (4) and (6) can be neglected. Eq. (7) is the condition of the *slope-defined resonance*. In optical fibers, it is common that the attenuation $\alpha < 10^{-6}$. Then, for the radiation wavelength $\lambda \sim 1.5\ \mu\text{m}$, effective refractive index $n_r \sim 1.5$, and radius $r \sim 50\ \mu\text{m}$, Eq. (7) is satisfied for $\gamma \gg 10^{-7}$, the experimental situation considered below.

For the slope-defined resonance, the transmission power, Eq. (6), is a linear combination of a constant and the real and imaginary parts of integral $\mathscr{P}(\Lambda) = -\int_0^\infty x^{-1/2}\exp(i\Lambda x + ix^3)dx$. This integral depends on the dimensionless wavelength shift $\Lambda = (96\pi^2 n_r^2 r^2 \lambda_q^{-5}\gamma^{-2})^{1/3}\Delta\lambda$, where, for convenience, the resonance wavelength, $\lambda_p = 2\pi n_r / \beta_p$, and wavelength shift, $\Delta\lambda = \lambda - \lambda_q$, are introduced. The plots of real and imaginary parts of function $\mathscr{P}(\Lambda)$, known as the generalized Airy function, are shown in Fig. 3. Similar to the ordinary Airy function, this function has asymmetric oscillations vanishing away from the principal peak. The dimensionless width of this peak, $\Lambda_0 \approx 5$, corresponds to the characteristic width of the principal spectral resonance:

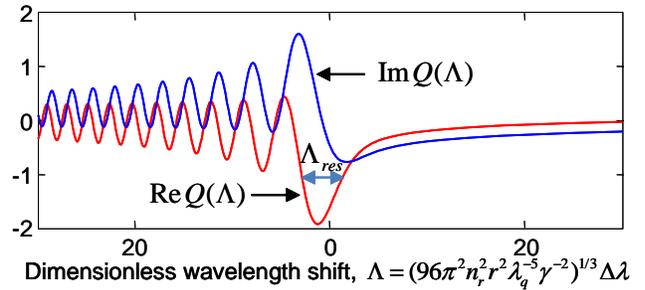

Fig. 3. Plots of the real and imaginary parts of the generalized Airy function $Q(\Lambda)$ as a function of dimensionless wavelength shift $\Lambda = (96\pi^2 n_r^2 r^2 \lambda_q^{-5}\gamma^{-2})^{1/3}\Delta\lambda$.

$$\Delta\lambda_0 \approx 0.5 n_r^{-2/3}\lambda^{5/3}\gamma^{2/3} r^{-2/3} \tag{8}$$

For small slopes $\gamma$, the value $\Delta\lambda_0$ decreases as $\gamma^{2/3}$ and determines the spectral resolution requested in the experiment. For example, for the optical fiber radius $r \sim 50\,\mu m$, wavelength $\lambda \sim 1.5\,\mu m$, refractive index $n_r \sim 1.5$, and fiber slope $\gamma \sim 10^{-5}$, we have $\Delta\lambda_0 \approx 23\,pm$. In this case, the identification of the resonance structure of Fig. 3 requests a pm wavelength resolution.

The behavior of conical modes in the slope-defined limit (and, more generally, for the negligible attenuation, $\alpha = 0$) can be better understood by introduction of the dimensionless propagation constant, $\overline{\Delta\beta}$, and axial coordinate $\overline{z}$:

$$\overline{\Delta\beta} = \Delta\beta / \Delta\beta_0, \quad \overline{z} = z / z_0,$$
$$\Delta\beta_0 = r^{-2/3}\beta^{1/3}\gamma^{2/3}, \quad z_0 = r^{1/3}\beta^{-2/3}\gamma^{-1/3}. \quad (9)$$

With these variables, the integral in Eq. (4) is simplified to

$$I(\overline{z},\overline{\Delta\beta}) = C\int_0^\infty \frac{d\overline{m}}{\overline{m}^{1/2}} \exp\!\left(i\Phi(\overline{\Delta\beta},\overline{z},\overline{m})\right),$$
$$\Phi(\overline{\Delta\beta},\overline{z},\overline{m}) = (2\overline{\Delta\beta}-\overline{z})\overline{m} - \frac{1}{3}\overline{m}^3 + \frac{\overline{z}^2}{4\overline{m}}, \quad (10)$$
$$C = \pi^{-1/2}(r\beta)^{-1/6}\gamma^{-1/3}.$$

In [10] it was shown that this integral can be transformed into a two-dimensional integral, which describes the characteristic field distribution near a caustic of the hyperbolic umbilic type $D_4^+$ in the configuration space with coordinates $\overline{\Delta\beta}$ and $\overline{z}$ [4,11]. Fig. 4(a) shows the surface plot of $|I(\overline{z},\overline{\Delta\beta})|$ which resembles this characteristic field distribution. However, in the considered situation, $\overline{\Delta\beta}$ is a fixed parameter rather than a coordinate in the configuration space. As the result, while the field distribution in Fig. 4(a) is not localized in plane $(\overline{z},\overline{\Delta\beta})$, it is localized for certain values of $\overline{\Delta\beta}$ along the coordinate $\overline{z}$. Obviously, the localization along the coordinate $\overline{z}$ means the full localization of the conical WGMs. The interesting features of the conical modes, which are observed in Fig. 4(a) are as follows. At large negative $\overline{\Delta\beta}$, the effect of slope can be neglected and the modes are localized along the cone axis due to interference effects, similar to the modes in a cylindrical microresonator [9]. An example of the WGM axial distribution for $\overline{\Delta\beta} = -2$ is given in Fig. 4(b), curve *1*. The behavior of conical modes for positive $\overline{\Delta\beta}$ is more complex. In this case, vanishing of the WGMs at positive $\overline{z}$ is explained by reflection at the narrower side of the cone from a turning point, similar to the conventional ray picture (Fig. 1(a)). For negative $\overline{z}$, most of the $\overline{\Delta\beta}$ values correspond to the slow decrease of the WGM amplitude, which is proportional to $\sim (-\overline{z})^{-1/4}$ for large $\overline{z}$ (see Appendix). This is the typical Airy function dependence near a turning point [3,4]. Curve *2* in Fig. 4(b) is an example of such WGM behavior for $\overline{\Delta\beta} = 0$. Crucially, it can be also found that, for the discrete sequence of dimensionless deviations of propagation constant, $\overline{\Delta\beta} = \overline{\Delta\beta}_n$,

$$\overline{\Delta\beta}_n = \frac{1}{2}\left(\frac{9\pi}{4} + 3\pi n\right)^{2/3}, \quad n = 0,1,2,..., \quad (11)$$

the complete destructive wave self-interference happens at the wider side of the cone. As the result, the conical modes experience the full localization. Derivation of Eq. (11) is given in the Appendix. Returning to the dimensional $\Delta\beta = \beta - \beta_q$ yields the quantization rule for the propagation constant of the determined localized conical states:

$$\beta_{qn} = \frac{q}{r} + \frac{1}{2}\left(\frac{\beta\gamma^2}{r^2}\right)^{1/3}\left(\frac{9\pi}{4} + 3\pi n\right)^{2/3} \quad (12)$$

where *n* is a positive integer and *q* is a large positive integer. Curves *3* and *4* in Fig. 4(b) are the conical mode distributions for $\overline{\Delta\beta} = \overline{\Delta\beta}_0$ and $\overline{\Delta\beta} = \overline{\Delta\beta}_2$, respectively. From Eq. (9), the characteristic size of the localized conical mode is

$$z_0 = r^{1/3}\beta^{-2/3}\gamma^{-1/3}. \quad (13)$$

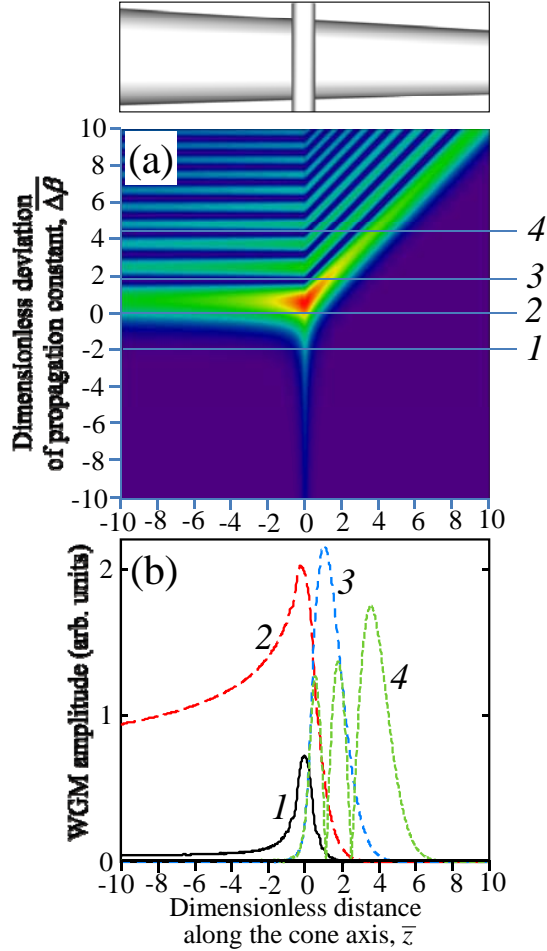

Fig. 4. (a) – A surface of conical WGM amplitude distribution along the dimensionless axial coordinate $\overline{z}$ vs. dimensionless deviation of propagation constant $\overline{\Delta\beta}$. (b) – WGM amplitude distributions for the values of … indicated by lines *1,2,3*, and *4* in (a). *1* – $\overline{\Delta\beta} = -2$, *2* – $\overline{\Delta\beta} = 0$, *3* – $\overline{\Delta\beta} = \overline{\Delta\beta}_0$, *4* – $\overline{\Delta\beta} = \overline{\Delta\beta}_2$. Inset: Illustration of a conical segment with an attached microfiber.

The $\gamma^{-1/3}$ dependence in Eq. (13) is very slow so that a conical resonator with an extremely small slope $\gamma$ can support strongly localized states. In fact, for the optical fiber with parameters $r \sim 50~\mu$m, $\lambda \sim 1.5~\mu$m, $n_r \sim 1.5$, and slope $\gamma \sim 10^{-5}$, we have $z_0 \approx 50~\mu$m.

## 3. Experiment

Experimentally, a 50 mm segment of an uncoated silica optical fiber with radius $r \approx 76~\mu$m was investigated. First, the radius variation of this fiber was accurately measured (Fig. 5). To this end, a biconical adiabatic fiber taper having the microfiber waist of a 1.3 $\mu$m diameter and 3 mm length was fabricated using the indirect $CO_2$ laser tapering method [14]. The ends of the taper were connected to the JDS Uniphase tunable laser source and detector. The wavelength resolution of this system was 3 pm. The microfiber was attached normally to the tested fiber where the WGMs were excited as illustrated in Fig. 1(b). The resonant transmission spectra were measured at points spaced by 2 mm along the tested fiber in the wavelength interval between 1535 nm and 1545nm. The coupling between the microfiber and the tested fiber was tuned to small values by shifting the microfiber/tested fiber contact point to a thicker part of the microfiber (see, e.g., samples of the measured spectra in Fig. 6(a) and (b)). To arrive at the plot of Fig. 5, the radius variation, $\Delta r$, was calculated from the shift of resonance positions, $\Delta\lambda$, as $\Delta r = \lambda\Delta\lambda/r$ [15].

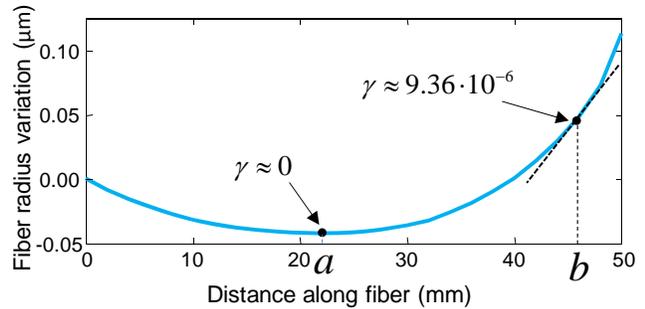

Fig. 5. Radius variation of a 50 mm optical fiber segment obtained experimentally with 2 mm measurement steps.

In order to experimentally verify the described theory, the transmission spectra at two positions of the tested fiber segment were examined. At the first position (point *a* in Fig. 5) the fiber slope was negligible and the corresponding transmission resonant spectrum shown in Fig. 6(a) was primarily determined by the WGM attenuation $\alpha$ [9]. At this position, the characteristic resonance width was smaller than the measurement resolution (see e.g., the magnified resonance sample in Fig. 6(c)) that allowed to estimate the attenuation constant as $\alpha < 10^{-6}$. Alternatively, at the second position (point $b$ in Fig. 5) the fiber slope found from the direct measurement presented in Fig. 5 gave $\gamma \approx 9.36 \cdot 10^{-6}$. This value, together with the estimate $\alpha < 10^{-6}$, satisfies the condition of the slope-defined resonances, Eq. (5). The latter suggests that the resonant transmission spectrum measured at point $b$ should exhibit the characteristic asymmetric oscillating behavior outlined in Fig. 3. The magnified structure of a separate resonance shown in Fig. 6(d) confirms this prediction. The experimental data in this figure was fitted by the linear combination of $\mathrm{Re}\,\mathscr{P}(\Lambda)$ and $\mathrm{Im}\,\mathscr{P}(\Lambda)$ assuming $r \approx 76\,\mu\mathrm{m}$, $\lambda = 1.54\,\mu\mathrm{m}$, and $n_r = 1.46$, which resulted in $\gamma \approx 9.354 \cdot 10^{-6}$ with better than 0.1% accuracy. This value is remarkably close to the directly-measured slope value $\gamma \approx 9.36 \cdot 10^{-6}$. Hence, the excellent agreement between the experimentally-obtained and theoretically-predicted shapes of resonances shown in Fig. 6(d) and also between the values of $\gamma$ found by independent measurements is observed.

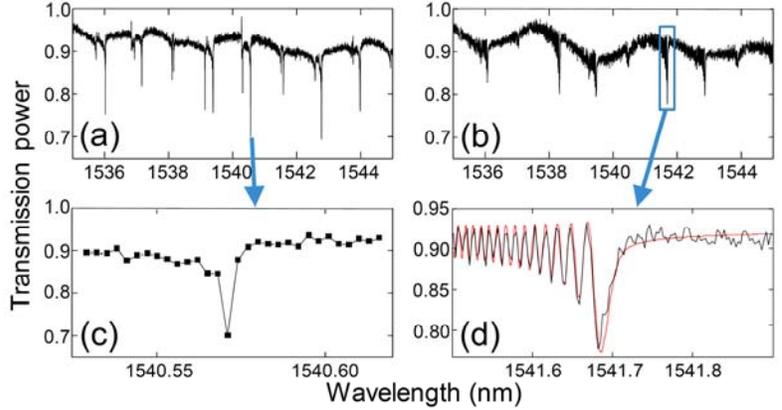

Fig. 6. (a) and (b) – resonant transmission spectra of the optical fiber at positions $a$ and $b$, respectively, shown in Fig. 5; (c) – plot (a) magnified near a single resonance; (d) – plot (b) magnified near a single resonance.

## 4. Summary and discussion

Thus, a small linear variation of an optical microcylinder radius leads to the appearance of strongly localized conical WGMs. These modes possess the discrete spectrum which is determined by a simple analytical expression, Eq. (12). The characteristic size of the localized conical modes grows very slowly with decreasing of the slope value so that a conical resonator with an extremely small slope $\gamma$ (e.g., a segment of an optical fiber) can support strongly localized states of light. The physical reason of the full WGM localization is the self-interference of the circulating beam. The reflection of the beam from the narrower side of the cone happens always and is similar to the reflection of classical rays propagating along the conical surface. The reflection of the beam from the wider side of the cone is caused by the destructive interference, which happens if the quantization rule, Eq. (12), is satisfied. The developed theory of conical modes predicts asymmetric oscillatory behavior of transmission spectrum which allows to determine the local change of fiber radius with unprecedented accuracy. Generally, an optical fiber with radius variation $\Delta r(z)$ can host different types of localized WGMs. The considered case $\Delta r(z) = \gamma z$ is related to the situation when the classical motion at the fiber surface is unbounded and, in particular, the circumference $(\varphi,0)$ is not a geodesic. Alternatively, for characteristic quadratic dependencies of $\Delta r(z) = -(z/z_0)^2$, this circumference is a stable geodesic which supports WGMs that are bounded along the axial direction by two caustics. These WGMs correspond to the conventional Lorentzian spectral resonances [1,2]. For other fiber profiles, more complex localized structures (e.g., WGM bottles [12]) can be present. It should be noted that the type of a resonance mode described in this paper is not limited to the special case of a constant conical slope. Deformation of the conical shape will result in appearance of other oscillatory spectral dependencies. Then, the terms depending on $\gamma$ in Eqs. (2), (3), (4), and (6) should be replaced by a more general function of $m$ determined by the equations of classical motion on the deformed surface. These more complex cases will be considered elsewhere.

The author is grateful to D. J. DiGiovanni and J. Fini for useful discussions and to Y. Dulashko for assisting in the experiments.

**Appendix: Derivation of the quantization rule for localized conical states**

It can be suggested from Fig. 4 that the conical states are localized for a discrete sequence of positive propagation constant displacements, $\overline{\Delta\beta}_n$, $n = 0,1,2,...$ For negative $\overline{z}$, the localization condition is defined by the equation

$$I(\overline{z},\overline{\Delta\beta}_n)\Big|_{\overline{z}\to-\infty} = 0. \tag{A1}$$

For large negative $\overline{z}$ and positive $\overline{\Delta\beta}$ the integral $I(\overline{z},\overline{\Delta\beta})$ can be calculated with the stationary phase method. In fact, expanding the expression for the stationary points $\overline{m}_{1,2}$ of this integral into series in small parameter $(-\overline{\Delta\beta}/\overline{z})^{1/2}$ yields:

$$\overline{m}_{1,2} = \left(\frac{-\overline{z}}{2}\right)^{1/2}\left[1 \mp \left(\frac{\overline{\Delta\beta}}{-\overline{z}}\right)^{1/2} - \frac{3}{2}\left(\frac{\overline{\Delta\beta}}{-\overline{z}}\right) \mp \left(\frac{\overline{\Delta\beta}}{-\overline{z}}\right)^{3/2} + ...\right]. \tag{A2}$$

For $\overline{m} = \overline{m}_{1,2}$

$$\Phi(\overline{\Delta\beta},\overline{z},\overline{m}_{1,2}) = \tfrac{1}{3}(-2\overline{z})^{3/2} + \overline{\Delta\beta}(-2\overline{z})^{1/2} \mp \tfrac{1}{3}(2\overline{\Delta\beta})^{3/2} + ...,$$

$$\frac{\partial\Phi(\overline{\Delta\beta},\overline{z},\overline{m})}{\partial\overline{m}}\Bigg|_{\overline{m}=\overline{m}_{1,2}} = 0, \tag{A3}$$

$$\frac{\partial^2\Phi(\overline{\Delta\beta},\overline{z},\overline{m})}{\partial\overline{m}^2}\Bigg|_{\overline{m}=\overline{m}_{1,2}} = \pm 2^{5/2}(\overline{\Delta\beta})^{1/2}.$$

As the result,

$$\int_0^\infty \frac{d\overline{m}}{\overline{m}^{1/2}}\exp\left(i\Phi(\overline{\Delta\beta},\overline{z},\overline{m})\right)\Bigg|_{\substack{\overline{z}\to-\infty \\ \overline{\Delta\beta}>0}} \approx \left(\frac{2}{-\overline{z}}\right)^{1/4}\exp\left[\tfrac{i}{3}(-2\overline{z})^{3/2} + i\overline{\Delta\beta}(-2\overline{z})^{1/2}\right] \times$$

$$\sum_\pm \int_{-\infty}^\infty d\overline{\Delta m}\exp\left[\mp\tfrac{i}{3}(2\overline{\Delta\beta})^{3/2} \pm 2^{3/2}i(\overline{\Delta\beta})^{1/2}\overline{\Delta m}^2\right] = \tag{A4}$$

$$\frac{(2\pi)^{1/2}\cos\left(\tfrac{1}{3}(2\overline{\Delta\beta})^{3/2} - \tfrac{\pi}{4}\right)}{(-\overline{\Delta\beta z})^{1/4}}\exp\left[\tfrac{i}{3}(-2\overline{z})^{3/2} + i\overline{\Delta\beta}(-2\overline{z})^{1/2}\right].$$

Thus, the condition for the localization of conical modes, Eq. (A1), is satisfied if

$$\cos\left(\tfrac{1}{3}(2\overline{\Delta\beta})^{3/2} - \tfrac{\pi}{4}\right) = 0. \tag{A5}$$

This equation is equivalent to the quantization rule of Eq. (11). The $\overline{z}$ dependence in Eq. (A4) resembles the asymptotic of the Airy function.